\pdfoutput=1

\documentclass[a4paper]{jpconf}

\usepackage{graphicx}
\usepackage{epsfig}

\usepackage{amsmath}
\makeatletter
\newcommand{\distas}[1]{\mathbin{\overset{#1}{\kern\z@\to}}}%
\newsavebox{\mybox}\newsavebox{\mysim}
\newcommand{\distras}[1]{%
  \savebox{\mybox}{\hbox{\kern3pt$\scriptstyle#1$\kern3pt}}%
  \savebox{\mysim}{\hbox{$\to$}}%
  \mathbin{\overset{#1}{\kern\z@\resizebox{\wd\mybox}{\ht\mysim}{$\to$}}}%
}
\makeatother

\begin{document}
\title{$\alpha$ clustering and flow in ultra-relativistic heavy-ion collisions}

\author{W Broniowski$^{1,2}$ and E Ruiz Arriola$^2$}

\address{$^1$ Institute of Physics, Jan Kochanowski University, 
25-406~Kielce, Poland}

\address{$^2$ H. Niewodnicza\'nski Institute of Nuclear Physics PAN, 31-342
Cracow, Poland}

\address{$^3$ Departamento de F\'{\i}sica At\'{o}mica, Molecular y Nuclear and
Instituto Carlos I de  F{\'\i}sica Te\'orica y Computacional, 
Universidad de Granada, E-18071 Granada, Spain}

\ead{Wojciech.Broniowski@ifj.edu.pl, earriola@ugr.es}

\begin{abstract}
We show how ultra-relativistic collisions of light nuclei with heavy
targets may be used to record snap-shots of the ground-state
configurations and reveal information on cluster correlations. The
development of collective flow in the formed fireball, which reflects
the geometric correlations in the initial state, is essential for the
method. As an illustration we analyze the $^{12}$C-$^{208}$Pb
collisions.
\end{abstract}

This talk is based on \cite{Broniowski:2013dia}, where the relevant
details can be found. We propose a novel way of investigating the
ground-state correlations in light nuclei, which provides a surprising
bridge between the lowest-energy nuclear structure and highest-energy
nuclear reactions, where collective flow of the fireball
develops. This flow transmutes the correlations in the initial state
(such as those due to the $\alpha$ clusters) into specific measurable
transverse-momentum asymmetries in the spectra of the produced
hadrons.

While the concept of the $\alpha$ clustering is more than 80 years old
\cite{gamow1931constitution} (for reviews see, e.g.,
\cite{brink1965alpha,freer2007clustered,ikeda2010clusters,beck2012clusters,%
Okolowicz:2012kv,Zarubin}), numerous issues still remain open. Even
the ground state structure of light nuclei, such as $^{12}$C, is a
topic of active research (see various contributions to these
proceedings). With this in mind, our method, inherently investigating
multiparticle correlations leading to collective effects (flow) may
provide a novel insight. We stress that since the clusterization
phenomenon concerns multi-particle correlations
(see, e.g.,~\cite{Viollier:1976ab}), it is accessible
directly only through observables which are many-body. Thus the
typically studied one-body quantities, such as excitation spectra of
the EM form factors, by definition cannot ``prove'' clusterization in
a direct manner.

\begin{figure}[tb]
\begin{center}
\epsfig{file=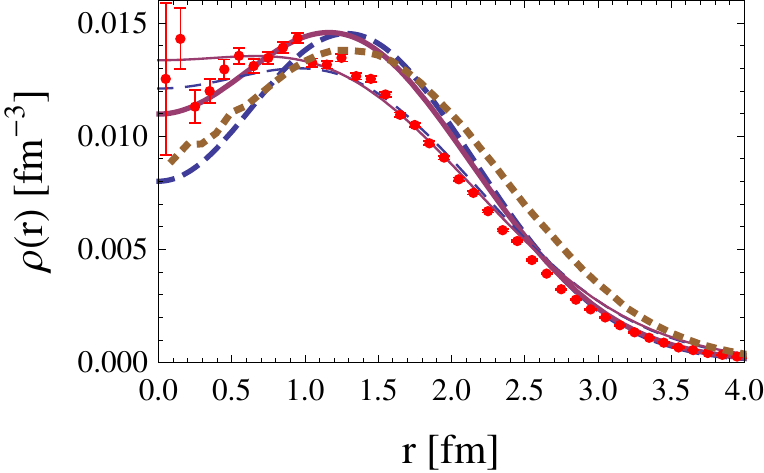,width=7.5cm,height=5cm}
\epsfig{file=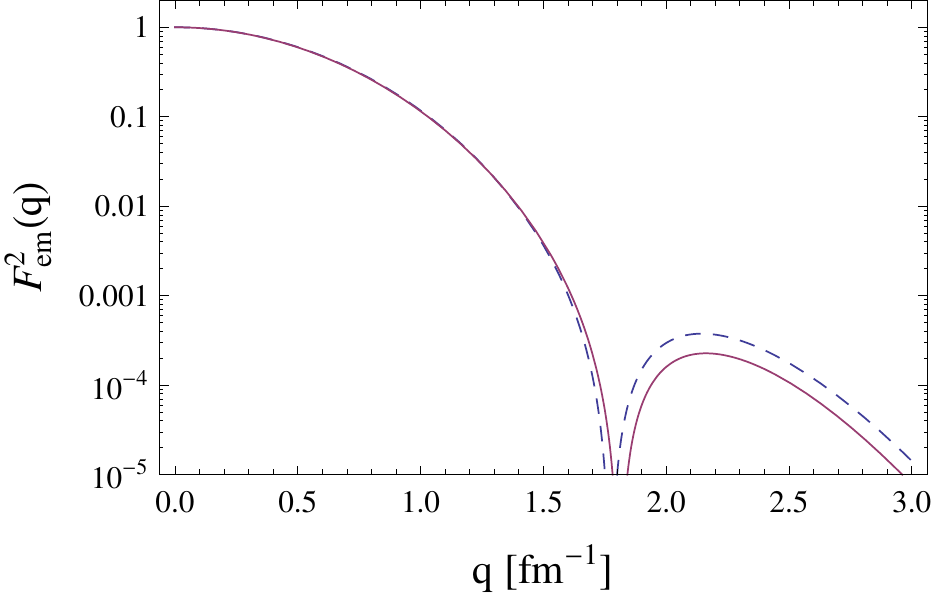,width=7.5cm,height=5cm}
\epsfig{file=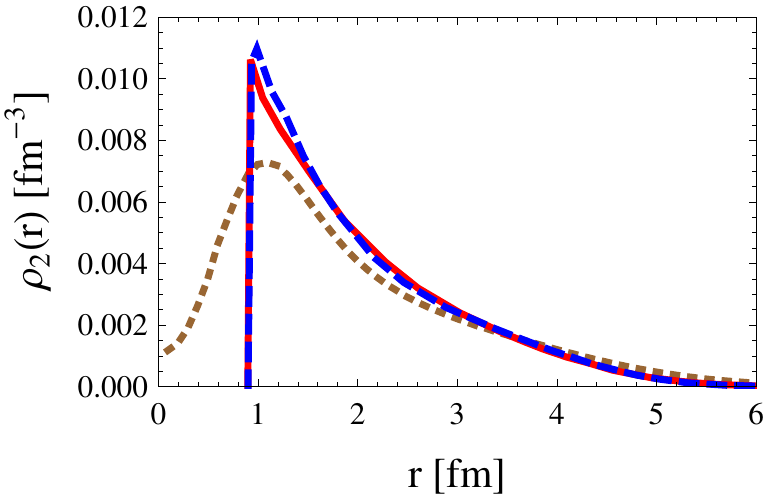,width=7.5cm,height=5cm}
\epsfig{file=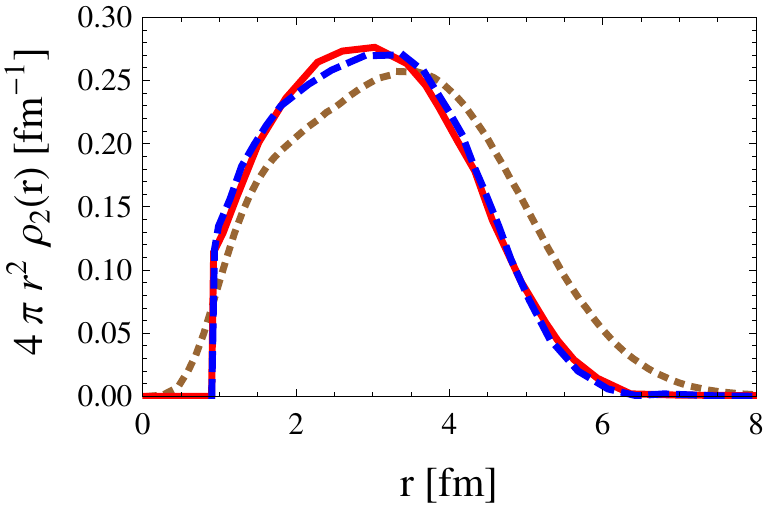,width=7.5cm,height=5cm}
\end{center}
\caption{\label{fig:dense} \small Upper left: Normalized one-particle
distributions in $^{12}C$.
The electric charge density $\rho^{\rm em}(r)/Z$ (thin lines) and the
corresponding distribution of the centers of nucleons $\rho_1(r)$
(thick lines) in ${}^{12}$C for the data~\cite{DeJager:1987qc} and
BEC~\cite{Funaki:2006gt} calculations (dashed lines),
FMD~\cite{Chernykh:2007zz} calculations (solid lines), and Jastrow
correlated wave function~\cite{Buendia:2004yt} (dotted line).  Upper
right: Charge Form factor $F^{\rm em}(q)$ in $^{12}C$ as a function of
the momentum transfer $q$ (in fm$^{-1}$) for BEC (dashed line) and FMD
calculations (solid line). Lower left: Normalized two particle distribution
$\rho_2(r)$ in $^{12}C$. We show our results for the fitted $\rho(r)$
in the FMD (thick line) and BEC cases (dashed line), and
compare to the Jastrow correlated wave function~\cite{Buendia:2004yt}
(dotted line). Lower right: The same as in lower left panel for the radial
distribution $4 \pi r^2 \rho_2(r)$.}
\end{figure}

To model the collision process, we specifically need the
distribution of centers of nucleons in ${}^{12}$C, which is nothing
but the ground state nuclear wave function squared, $|\Psi_A(\vec x_1
, \dots, \vec x_A)|^2$. While it would be best to incorporate
realistic calculations (see,
e.g.,~\cite{Buendia:2004yt,Wiringa:2013ala}), in
Ref.~\cite{Broniowski:2013dia} we have applied a simple and practical procedure
with $\alpha$-clustered (or unclustered for comparison) random
distributions.  In the
$\alpha$-clustered case we randomly generate positions of the 12
nucleons, 4 in each cluster of a Gaussian shape and size $r_\alpha$. The
centers of the
clusters are placed in an equilateral triangle of side length $l$. 
The short-distance NN repulsion is incorporated by
precluding the centers of each pair of nucleons to be closer than the
expulsion distance of 0.9~fm~\cite{Broniowski:2010jd}.
The parameters $l$ and $r_\alpha$ are optimized such
that the one particle density $\rho(r)$ of BEC~\cite{Funaki:2006gt}
or FMD~\cite{Chernykh:2007zz} calculations are accurately reproduced
(standard unfolding the proton charge density 
from the charge distribution $\rho^{\rm em}(r)$ is
necessary), see Fig.~\ref{fig:dense}.  Note a large central depletion
in the distributions, originating from the separation of the $\alpha$
clusters arranged in the triangular configuration. Besides, a fair
reproduction of two particle densities $\rho_2(r)$ from multiclustered
Jastrow correlated calculations~\cite{Buendia:2004yt} is observed.
The radial distribution $4\pi r^2\rho_2(r)$ peaks at the size of the
triangle, $l \simeq 3~{\rm fm}$.
Thus, we deal with realistic nuclear distributions.

\begin{figure}[b]
\begin{center}
\includegraphics[width=14pc]{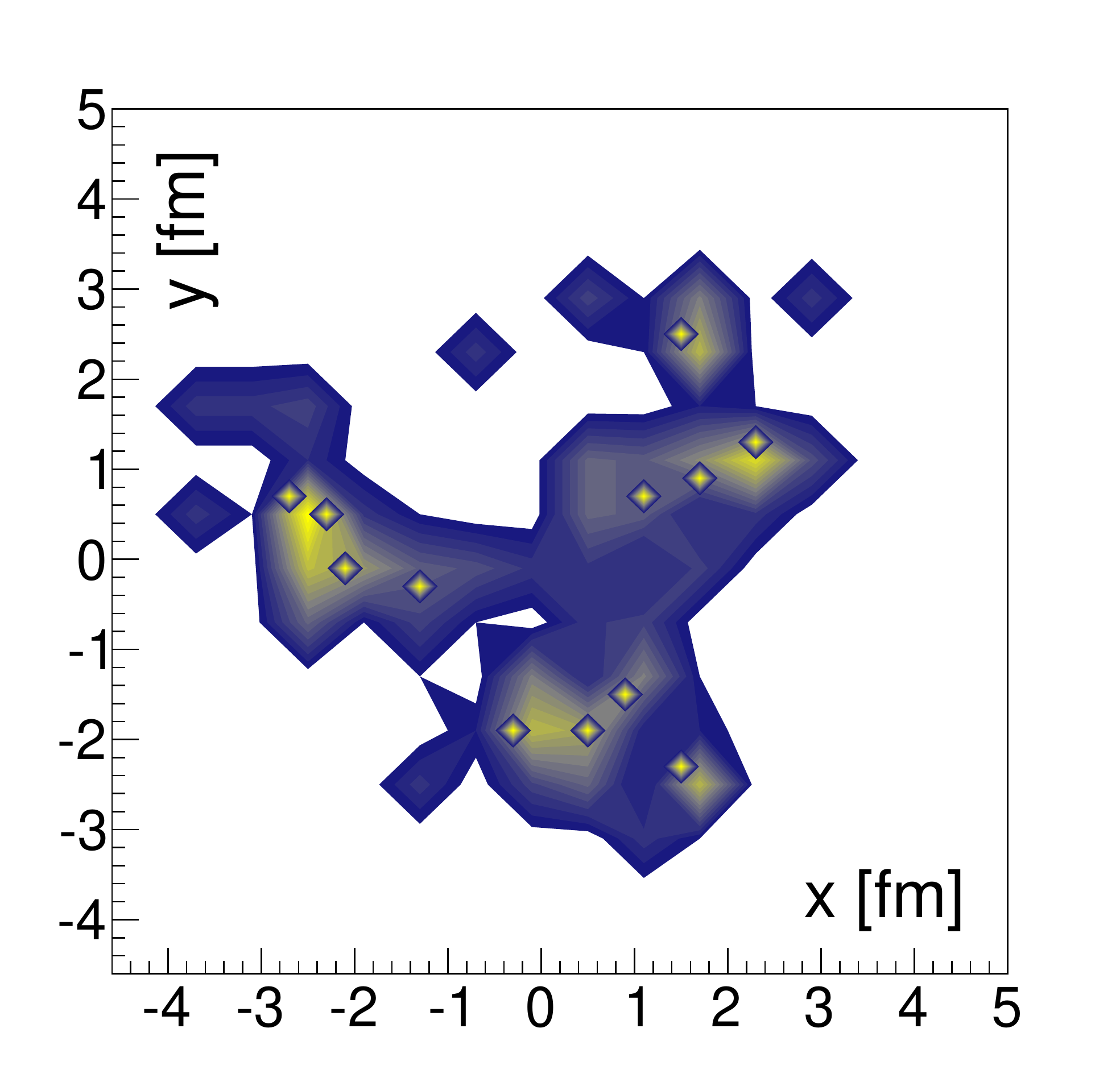}\hspace{2pc}%
\begin{minipage}[b]{14pc}\caption{\label{fig:fire} \small Fireball created in a
$^{12}$C-$^{208}$Pb collision, viewed in the transverse plane. The original
location of the nucleons in $^{12}$C is indicated by small diamonds. The
fireball is formed from the collisions of the projectile and target nucleons
within the Glauber model. The triangular arrangement of the three $\alpha$
clusters is reflected in the triangular shape of the fireball, warped to some
extent with fluctuations.}
\end{minipage}
\end{center}
\end{figure}

\begin{figure}[tb]
\begin{center}
\begin{minipage}{14pc}
\includegraphics[width=14pc]{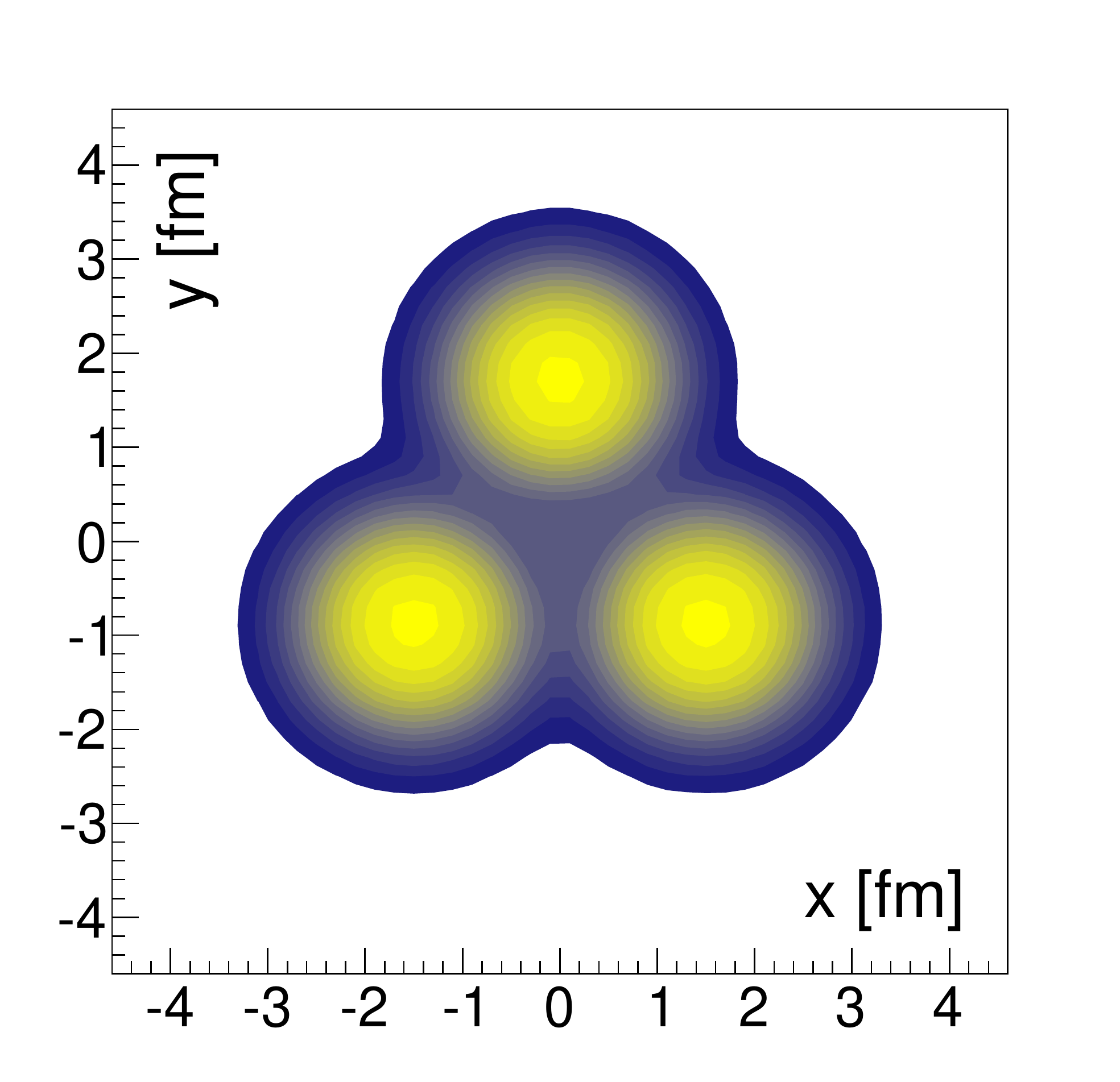}
\caption{\label{fig:intr0} \small Average intrinsic triangular distribution of
$^{12}$C
projected on the transverse plane.}
\end{minipage}\hspace{2pc}%
\begin{minipage}{14pc}
\includegraphics[width=14pc]{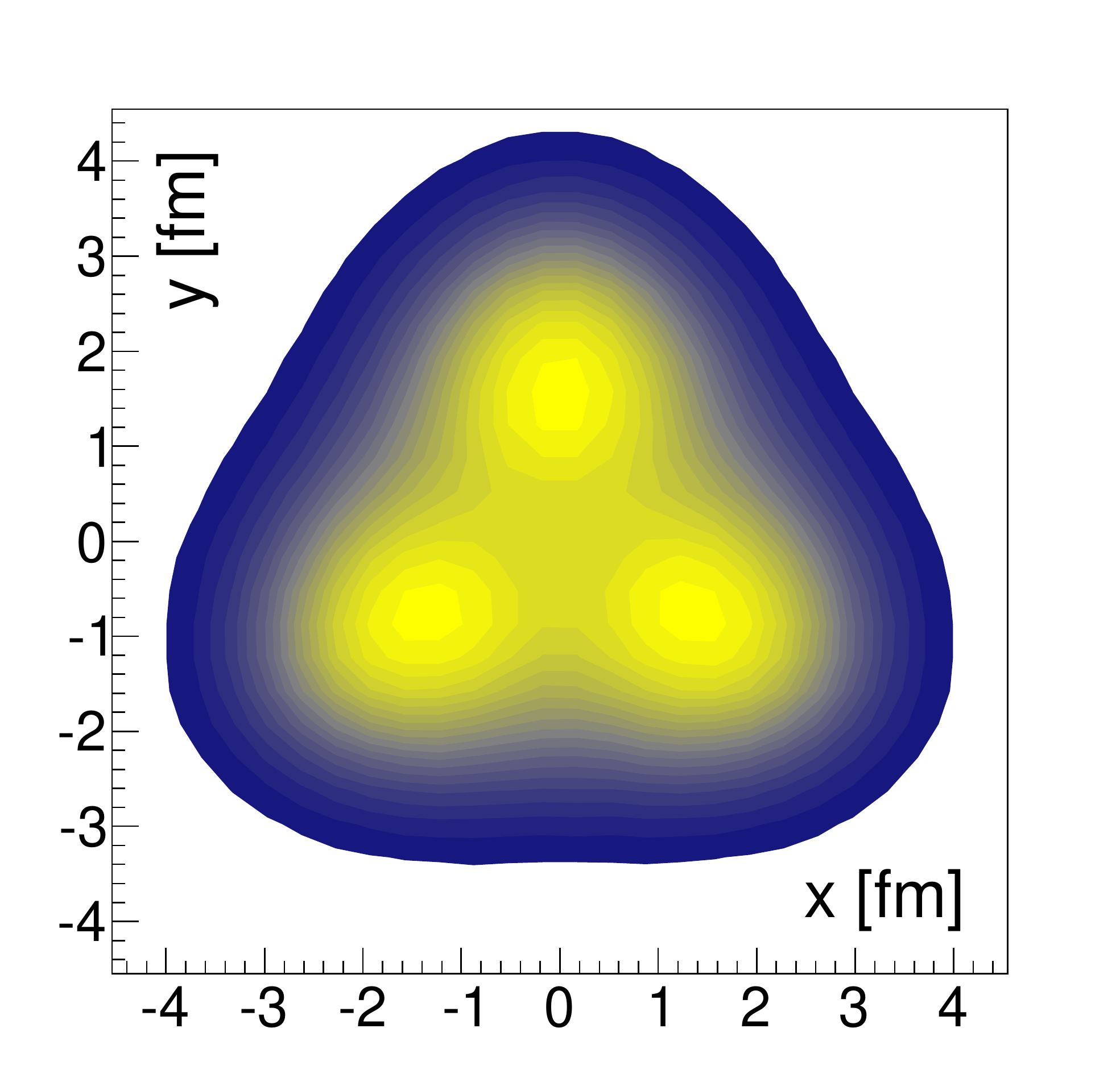}
\caption{\label{fig:intr} \small Corresponding average intrinsic transverse
density of
the fireball created in a flat-on $^{12}$C-$^{208}$Pb collision.}
\end{minipage}
\end{center}
\end{figure}

The essence of our approach is as follows: $\alpha$ clusters lead to
substantial {\em intrinsic} deformation of nucleon distributions in
light nuclei. Since the time of the reaction in ultrarelativistic
collisions is much shorter from any characteristic time of the nuclear
structure (the Lorentz contraction factor for the colliding nuclei
reaches 100 at RHIC!), an instant snapshot of the frozen light nucleus
configuration is made, revealing the lumpy structure when
present. Consequently, when a geometrically deformed nucleus hits a
large target at almost the speed of light, the created fireball in the
transverse plane inherits the shape of the light nucleus
(cf.~Fig.~\ref{fig:fire}-\ref{fig:intr}).  With a large target, the
created fireball is abundant enough to evolve collectively, in full
analogy to ultrarelativistic collisions of two heavy nuclei as studied
recently in colliders (RHIC, LHC) or fixed-target experiments (SPS).
Due to the initial fireball deformation, a deformed flow pattern
develops, leading to azimuthal asymmetry in the transverse momentum
distributions of the hadrons (mostly pions) produced in the
collision. This asymmetry can be analyzed and measured event-by-event
through well-established methods%
~\cite{Ollitrault:1992bk,Borghini:2001vi,Voloshin:2008dg}.
Here we focus on the promising $^{12}$C+$^{208}$Pb system, as the
$^{12}$C has a large intrinsic triangularity
\cite{KanadaEn'yo:2006ze,Chernykh:2007zz}, resulting in large triangular flow,
increasing strongly with the multiplicity of the produced hadrons
\cite{Broniowski:2013dia}. 

In our study we apply
GLISSANDO~\cite{Broniowski:2007nz,Rybczynski:2013yba} to model the
early phase of the collision within the Glauber Monte Carlo approach.
The eccentricity parameters $\epsilon_n$ are convenient measures of
the harmonic components of the intrinsic deformation. They are defined
for each collision (event) as the coefficients of the Fourier
decomposition of the distribution in the transverse plane,
\begin{eqnarray} 
\epsilon_n e^{i n \Phi_n} = \frac{\sum_j w_j \rho_j^n e^{i n
\phi_j}}{\sum_j w_j \rho_j^n},
\end{eqnarray}
where $j$ labels the {\em sources} in the event, $\rho_j$ is the transverse
position of the source, $w_j$ its weight, $n$ indicates the rank,
and, finally, $\Phi_n$ is the principal axis angle in the event.
The $n=2$ deformation is referred to as {\em ellipticity}, and $n=3$ as
{\em triangularity}. The notion of the {\em source} is used in the
Glauber-model sense, and indicates the wounded nucleons~\cite{Bialas:1976ed} or
binary collisions.

\begin{figure}[tb]
\begin{center}
\includegraphics[width=14pc]{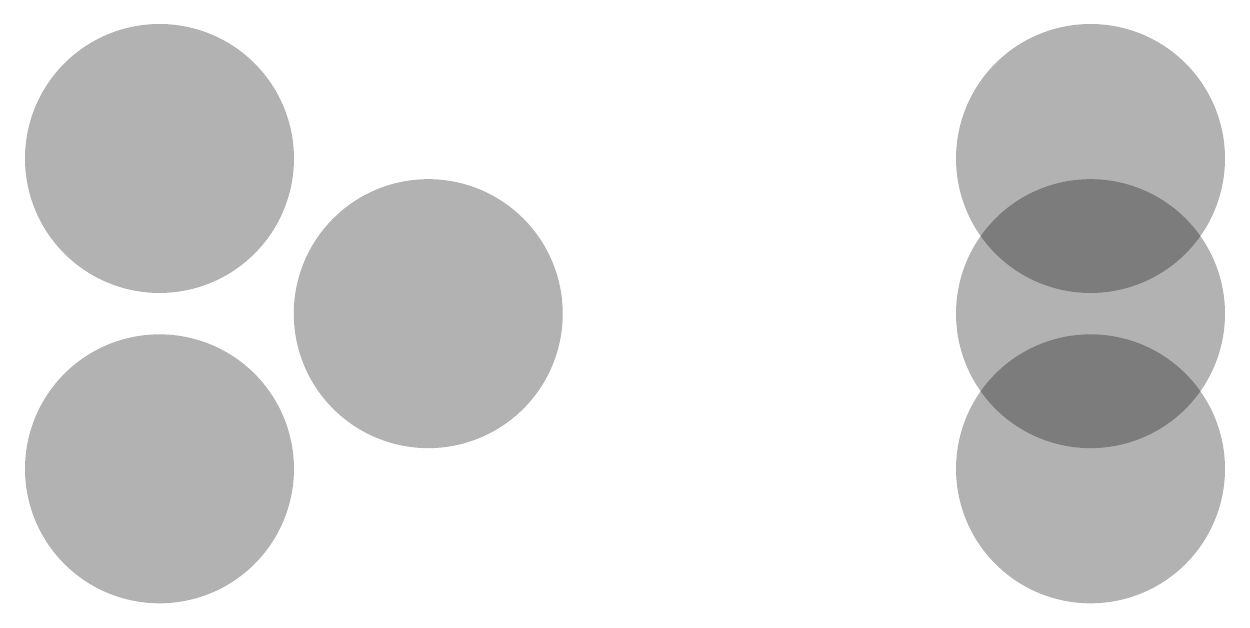}\hspace{2pc}%
\begin{minipage}[b]{14pc}\caption{\label{fig:flat} \small Flat-on (left) and
side-wise (right) orientations of the $^{12}$C nucleus in collisions
with $^{208}$Pb. The
former leads to a larger multiplicity, larger triangularity, and smaller
ellipticity, while the latter, on the opposite, has smaller multiplicity,
smaller triangularity, and larger ellipticity.}
\end{minipage}
\end{center}
\end{figure}

The collective dynamics of the intermediate evolution of an
ultra-relativistic nuclear reactions is expected to be properly
modeled with hydrodynamics (for recent reviews see,
e.g.,~\cite{Heinz:2013th,Gale:2013da} and references therein) or
transport models~\cite{Lin:2004en}.  These methods lead to an
event-by-event transmutation of the initial-state anisotropies,
quantified with the $\epsilon_n$ coefficients, into the harmonic flow
coefficients of the transverse momentum distributions of the produced
particles, known as
$v_n$~\cite{Ollitrault:1992bk,Borghini:2001vi,Voloshin:2008dg}.
Since it is well established that the proportionality $v_n \sim
\epsilon_n$ holds (see, e.g., a recent analysis~\cite{Bzdak:2013rya}), one may
gain the relevant information on the physically relevant $v_n$ coefficients by
studying the eccentricities $\epsilon_n$, which is way simpler, as no costly
hydrodynamic or transport studies are needed (such full-fledged studies are
underway).

Before showing our key results (here shown for the distributions
reproducing the BEC case), let us present a very specific and
crucial for understanding correlation between the ``geometry'' and the
multiplicity of particles produced in the collision (which is proportional to
the number of the wounded nucleons, $N_w$). The orientation of the ``little
triangle'' describing the $^{12}$C nucleus with respect to the transverse plane
is random in each event; sometimes the collision is flat-on, sometimes
side-wise, or assumes any intermediate angle (cf.~Fig.~\ref{fig:flat}). In the
flat-on case the damage created by the projectile in the target is largest, as
the geometric cross section is highest (at ultra-relativistic energies the
projectile nucleon wounds everything in its straight path!). At the same time,
triangularity is highest and the ellipticity is lowest.
For the side-wise orientation, the effects are opposite.
Thus we find positive correlation between multiplicity and triangularity, and
negative correlations between multiplicity and ellipticity.

\begin{figure}[tb]
\begin{center}
\includegraphics[width=14pc]{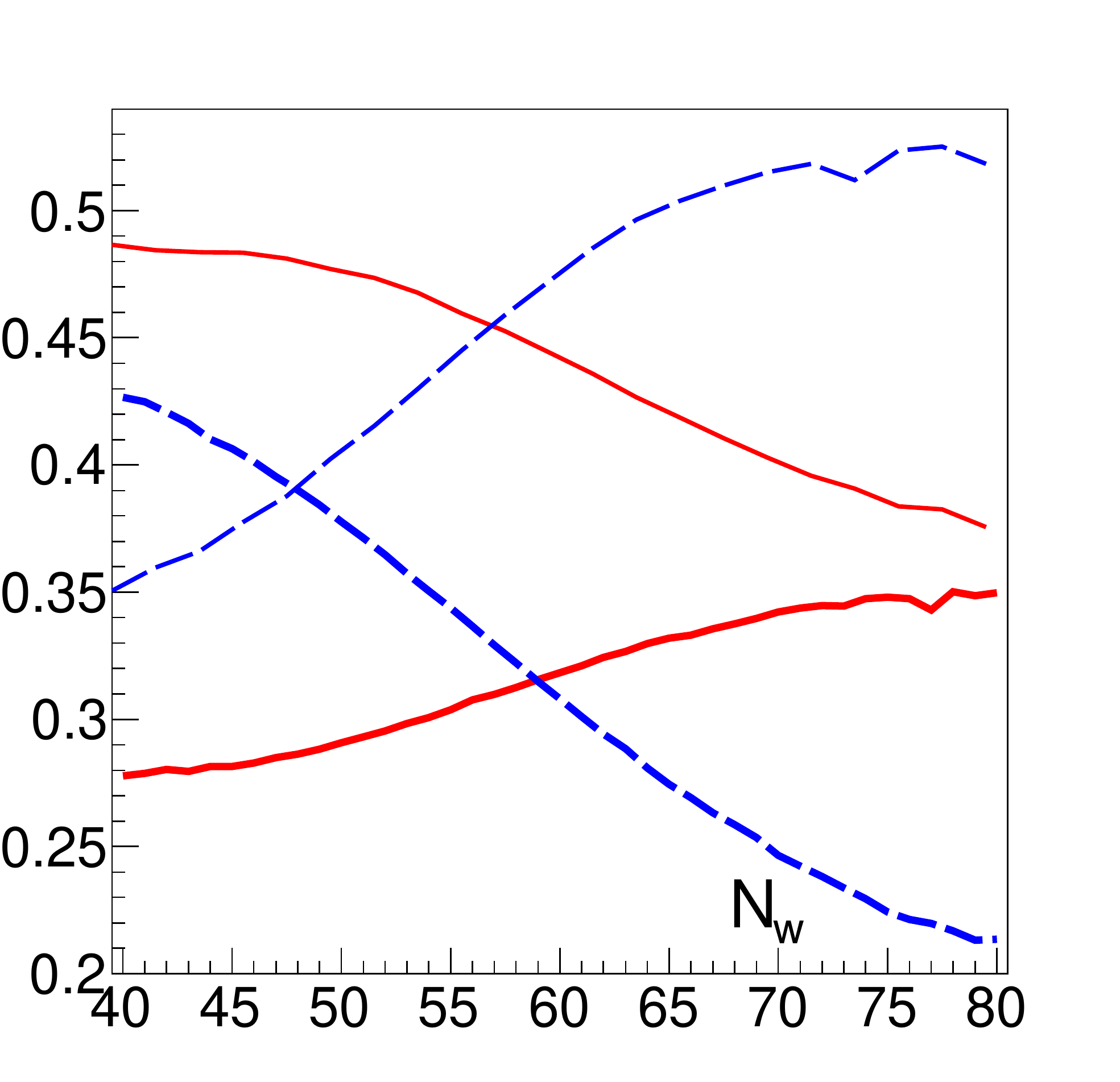} \includegraphics[width=14pc]{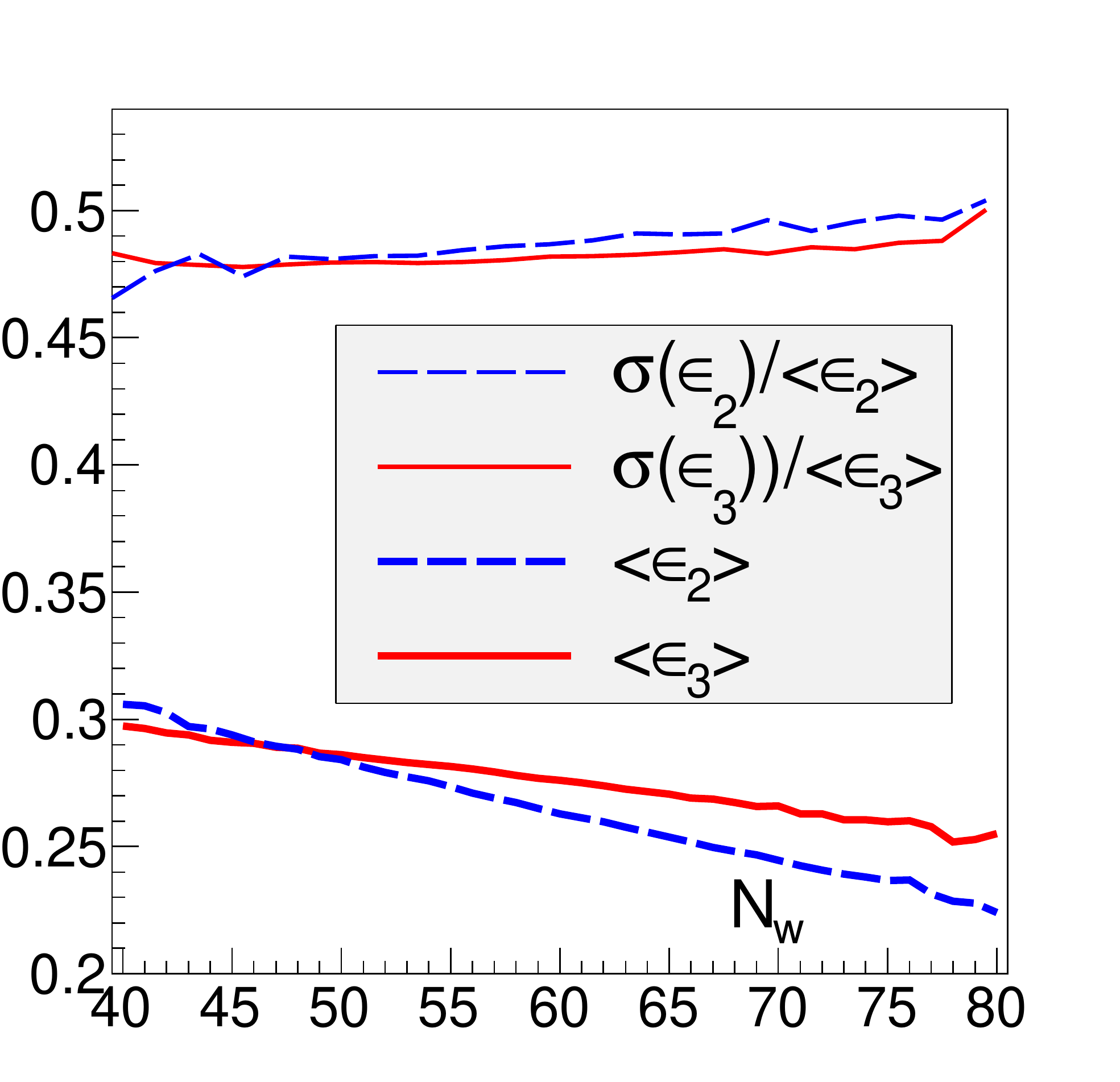}
\end{center}
\caption{\label{fig:eps} \small Event-by-event average ellipticity and
triangularity of the fireball, as well as their scaled standard deviations,
plotted as functions of the number of wounded nucleons~\cite{Bialas:1976ed},
$N_w$. Left: the case where the $^{12}$C nucleus is clustered according to the
BEC model. Right: uniform distribution (no clusters). SPS energies
($\sigma_{NN}^{\rm inel}=32$~mb), mixed
Glauber model~\cite{Broniowski:2013dia}.}
\end{figure}

This simple
quantitative effect is clearly seen in actual simulations with GLISSANDO,
presented in the left panel of Fig.~\ref{fig:eps}. The displayed growth of
the event-averaged $\epsilon_3$ and the falloff of the event-averaged
$\epsilon_2$ are the advocated signatures of the $\alpha$ clusterization in
$^{12}$C. The case of the unclustered $^{12}$C (i.e., with the nucleons
distributed uniformly but with exactly the same one-body density as in the
clustered case) is shown in the right panel of Fig.~\ref{fig:eps}. We note a
very similar behavior of $\langle \epsilon_2 \rangle$ and $\langle \epsilon_3
\rangle$ as functions of $N_w$. The fact that these are non-zero is due entirely
to fluctuations~\cite{Bzdak:2013rya} of the uniform system with a finite number
of sources. In addition, we show the scaled standard deviations of the
event-by-event
values of $\epsilon_3$ and $\epsilon_2$. These measures are also sensitive to
clusterization. 

We note that the presented analysis is similar in spirit to the studies of the
d-A collisions~\cite{Bozek:2011if} and
$^3$H/$^3$He-A collisions~\cite{Sickles:2013mua} (to be investigated
experimentally at RHIC), however, the collective effects in a larger system,
such as in $^{12}$C-$^{208}$Pb presented here, are expected to be much stronger,
hence the shape-flow transmutation should be more visible. In
particular, the separation of the geometric component of the flow from the
non-flow effects due to fluctuations should be easier.

In summary, our method,
which links the cluster features of lowest-energy nuclear structure physics with
flow phenomena known from the highest-energy nuclear collisions,
offers a novel possibility to study geometric correlations in the ground
state. Viewed from the opposite direction, 
a detailed knowledge of the clustered nuclear
wave functions (obtained, e.g., from the ab initio numerical analyses)
will help to place important constraints on the models
of the fireball evolution and thus
gain information on the properties of the quark-gluon plasma.

\ack

This research was supported by the Polish National Science Centre,
grants DEC-2012/05/B/\-ST2/\-02528 and DEC-2012/06/A/ST2/00390, Spanish
DGI (grant FIS2011-24149) and Junta de Andaluc\'{\i}a (grant FQM225).

\section*{References}

\bibliographystyle{iopart-num}
\bibliography{clusters,hydr,adds}

\end{document}